\documentclass[aps,prb,superscriptaddress,amsmath,amssymb,aps,twocolumn,]{revtex4-2}
\usepackage{graphicx}% Include figure files
\usepackage{dcolumn}% Align table columns on decimal point
\usepackage{bm}% bold math
\usepackage{latexsym,epsfig}
\usepackage{graphicx}
\usepackage{verbatim}
\usepackage{comment}
\usepackage{amsmath}
\usepackage{amssymb}
\usepackage{stmaryrd}
\usepackage{color}
\usepackage{epstopdf}
\usepackage{grffile}
\usepackage[normalem]{ulem}
\usepackage{float}
\usepackage{tikz}
\usetikzlibrary{quantikz}
\DeclareGraphicsExtensions{.eps}

\newcommand{\beq}{\begin{equation}}
\newcommand{\eeq}{\end{equation}}
\newcommand{\bea}{\begin{eqnarray}}
\newcommand{\eea}{\end{eqnarray}}
\newcommand{\ben}{\begin{eqnarray*}}
\newcommand{\een}{\end{eqnarray*}}
\newcommand{\bfig}{\begin{figure}}
\newcommand{\efig}{\end{figure}}

\usepackage{hyperref}
\hypersetup{
    colorlinks=true,      
    urlcolor=blue,
    citecolor=blue,
    linkcolor=blue
}

\begin{document}
\title{Anomalous slow-down of the bound state dynamics in a non-locally coupled quantum circuit}

\author{Biswajit Paul}
\affiliation{School of Physical Sciences, National Institute of Science Education and Research, Jatni 752050, India}
\affiliation{Homi Bhabha National Institute, Training School Complex, Anushaktinagar, Mumbai 400094, India}

\author{Suman Mondal} 
\affiliation{Max Planck Institute for the Physics of Complex Systems, Nöthnitzer Strasse 38, 01187 Dresden, Germany}

\author{Tapan Mishra}
\email{mishratapan@gmail.com}
\affiliation{School of Physical Sciences, National Institute of Science Education and Research, Jatni 752050, India}
\affiliation{Homi Bhabha National Institute, Training School Complex, Anushaktinagar, Mumbai 400094, India}

% \author{Biswajit Paul$^{1,2}$, Suman Mondal$^{3}$, and Tapan Mishra$^{1,2}$}
% \affiliation{
% $^1$School of Physical Sciences, National Institute of Science Education and Research, Jatni 752050, India\\
% $^2$Homi Bhabha National Institute, Training School Complex, Anushaktinagar, Mumbai 400094, India\\
% $^3$Max Planck Institute for the Physics of Complex Systems, Nöthnitzer Strasse 38, 01187 Dresden, Germany}

\date{\today}

\begin{abstract}

Additional hopping channels in a tight-binding lattice is known to introduce faster dynamics of a quantum mechanical particle. However, we show that in the case of a repulsively bound state, the dynamics becomes abnormally slow when next-nearest neighbor (NNN) hopping is allowed for the particles. We show that such slowing down occurs for some magic strength of the NNN hopping at which the bound state band exhibits a quasi-flatband feature. We reveal this anomalous dynamical behavior by analyzing the quench dynamics of two nearest neighbor (NN) spin excitations (magnons) on a ferromagnetic chain by allowing both NN and NNN couplings. By implementing digital quantum computing simulations on a NISQ device, we obtain such non-trivial signatures and complement the results with exact numerical calculations. Moreover, through perturbative arguments, we reveal that the slowing down is due to the destructive interference between different paths associated to the bound state dynamics.

\end{abstract}

\maketitle
{\em Introduction.-} 
Non-equilibrium dynamics of interacting quantum systems are known to reveal novel scenarios which are completely different from the systems at equilibrium. Recent theoretical developments complemented by experimental observation using various quantum simulators have enriched our understanding of several interesting phenomena in systems out of equilibrium~\cite{rev1, rev2, rev3_1, rev3_2, rev4, rev5, rev6, rev7, rev8, rev9, rev11, rev12, rev10, RevBloch2008}. This includes many-body localization~\cite{Altman2019, Sierant_2025}, quantum scars~\cite{qscar1_1, qscar1_2,qscar2_1,qscar2_2,qscar2_3, Bernien2017, Wei2019, Choi2019, Moudgalya_2022, Chandran2023}, Hilbert space fragmentation~\cite{ Scherg2021,Kohlert2023, Somsubhra2023, Somsubhra2024, paul2025}, time crystal~\cite{tc1_1, tc1_2,tc2_1, tc2_2, tc2_3,tc3_1,tc3_2,tc3_3,tc4_1,tc4_2,tc4_3}, dynamical phase transitions~\cite{DQPT2013, Heyl_2018}, robust topological edge excitations~\cite{topo1_1, topo1_2, topo1_3,topo2_1,topo2_2,topo3_1,topo3_2,topo4_1,topo4_2,topo4_3, Jangjan2020, Wintersperger2020, Nur2023, Nur2025}, etc. In this context, quench dynamics of an initial quantum state with respect to a final Hamiltonian constitutes an important protocol for driving a system away from equilibrium. While the quenching of a many-body state often faces experimental challenges, the quench dynamics of a few-particle system or quantum walk (QW) offers a unique and simplest platform to understand the dynamical behavior of interacting particles. Recent years have seen a growing interest in this direction, revealing numerous interesting phenomena, proving QW one of the most suitable approachs for studying fundamental physics and technological applications~\cite{Knight2003, Shenvi2003,Childs2004, Pathak2007, Tulsi2008, Childs2009, Lovett2010, Peruzzo2010, stefanak_2011, Lahini2012, Andrew2013, Qiang2016, Rigovacca2016, Caruso_2014, Preiss2015, Chattaraj2016, Costa2019, Mondal2020, yan2019, Gong2021, Giri2021, Cai2021, Giri2022, Ostahie2023, giri2023, giri2024, Paul2024, maity2024, Tao2024, Camacho2025, rev_QW}. 

One notable observation that has emerged from the QW of interacting particles is of the repulsively bound pairs. Although the existence of such a non-trivial entity was discussed in the context of Hubbard models and spin chains~\cite{Bethe1931, Hubbard1963, Yang1989, Silberglitt1970, Tonegawa1970}, their experimental observation remained challenging until the first concrete evidence using ultracold atoms in optical lattices~\cite{Winkler2006} and recently in condensed matter materials~\cite{CorinnaKollath2024}. However, given the feasibility of obtaining the signatures in the few particle quench dynamics ~\cite{Ganahl2012} and subsequent experimental advancements have resulted in the observation of repulsively bound pairs in various platforms ~\cite{Bloch2013,Corrielli2013,Florian2023}. 
%ir approach, the QW the and observed in experimental observation remained . On the other hand, as the high energy states lying above the band of scattering state, their observation in conventional solid state systems were challenging~\cite{}. \biswajit{However, their recent observation in ultracold $^{87}$Rb atoms in optical lattice and subsequent theoretical proposals~\cite{Ganahl2012} for their observation in quench dynamics~\cite{Bloch2013} have prompted for the observation of repulsively bound pairs in the QW using various set ups such as ultracold atoms in optical lattices~\cite{Winkler2006, Bloch2013}, photonic and optical systems~\cite{Corrielli2013}, superconducting circuits~\cite{Google2022} and array of trapped ions etc~\cite{Florian2023}.} 
Going beyond the conventional two-particle dynamics, a recent study using Google's superconducting quantum circuit has revealed the remarkable signatures of bound states of up to five particles in the QW ~\cite{Google2022} opening up directions to further explore the nature of these non-trivial bound states~\cite{Surace2024, Hudomal2024}. 
%This has also spurred interest in the simulation of QW of interacting particles using appropriate quantum algorithms~\cite{}. 

The observation central to all these findings is the slower spreading of the wavefunction in the QW due to the reduced effective hopping of the bound state compared to that of the individual non-interacting constituents. 
%This slower spreading is due to the formation of repulsively bound state which evolves with an effective hopping much smaller than the hopping of the individual constituents. 
The velocity of propagation of the composite particle, although suppressed, is strongly dependent on the dispersion of the bound state in the energy spectrum of the few-body problem. Depending on whether the interaction (hopping) is stronger (weaker), the bound state band becomes less dispersive, resulting in slower dynamics of the bound state. On the contrary, with increase in hopping strength or with any additional hopping paths in the system, the bound state band is expected to become more dispersive and the dynamics faster as is expected from the dynamics of non-interacting particles. 
%This preceding conclusion is true for a single particle QW. 
However, in this Letter, we argue that in the case of a repulsively bound state of particles, additional hopping channel in the system results in a robust slowing down of the dynamics. We demonstrate this remarkable phenomenon by considering a pair of NN spin excitations in a ferromagnetic spin-1/2 chain with NN and NNN couplings. We show that for strong NN longitudinal coupling and in the absence of NNN transverse coupling, the two spin excitations (magnons) form a bound magnon pair exhibiting slower dynamics compared to the individual magnon dynamics. However, with the onset of the NNN transverse coupling, a drastic reduction in the spreading of wavefunction of the magnon bound state appears in the QW at a critical NNN transverse coupling strength. Eventually, with further increase in the strength of the NNN transverse coupling, a faster spreading returns in the QW as expected. In addition to this, we show that the spreading is fully suppressed for some critical values of NNN transverse coupling strength, if the magnons are initialized at one of the edges of the chain. Such a non-trivial slowing down of the magnon bound state provides a scenario of intricate interplay between various couplings in the system.

{\em Model.-}
The Hamiltonian that governs the dynamics of the spin excitations in our study is the one-dimensional spin-$\frac{1}{2}$ chain is given as
\begin{align}
    \hat{H} = &-\frac{J_{1}}{2}\sum_{\langle i, j\rangle} (\hat{\sigma}_{i}^x \hat{\sigma}_{j}^x+\hat{\sigma}_{i}^y \hat{\sigma}_{j}^y)
    -\frac{J_{2}}{2}\sum_{\langle\langle i, j\rangle\rangle} (\hat{\sigma}_{i}^x \hat{\sigma}_{j}^x+\hat{\sigma}_{i}^y \hat{\sigma}_{j}^y)\nonumber\\ 
    &+\frac{J_{z}}{4}\sum_j (1+\hat{\sigma}_j^z)(1+\hat{\sigma}_{j+1}^z) 
    \label{eq:ham}.
\end{align}
Here, $\hat{\sigma}_j^x, \hat{\sigma}_j^y$ and $\hat{\sigma}_j^z$ represent the Pauli matrices and $J_{z}$ is the longitudinal NN coupling between the spin excitations. $\langle i,j\rangle$ and $\langle\langle i,j \rangle\rangle$ denote the coupling between NN and NNN spins with coupling strengths $J_1$ and $J_2$ respectively. The last term in the Hamiltonian ensures that the system is not domain-wall conserving. Note that this model can be mapped to a hardcore bosonic model featuring NN and NNN hoppings and NN interaction, using the Holstein–Primakoff transformation~\cite{HolsteinPrimakoff}. For our studies, we define all the physical quantities in units of $J_1$ by setting it equal to one.

The model above, in the absence of $J_2$ reduces to the well-known XXZ model with a transverse field. In this context, the dynamics of two NN spin excitations on a ferromagnetic chain has been experimentally observed using ultracold atoms in optical lattices~\cite{Bloch2013}. It has been shown that in the limit of $J_z>>J_1$, a magnon bound state at NN sites is formed exhibiting much slower dynamics compared to that of a single magnon. Recently, a similar phenomenon has been observed in the form of the bound states of microwave photons in an array of superconducting circuits~\cite{Google2022}. In the following, we explore the effect of $J_2$ on the bound state dynamics. 

%Note that the model shown above can also be mapped to system of NN interacting hardcore bosons on a one dimensional lattice with both NN and NNN hoppings. 

The quantum state at a time $t \neq 0$ can be determined from the solution of the time-dependent Schrodinger equation $|\Psi(t)\rangle = e^{-i\hat{H}t}|\Psi(0)\rangle$, where $|\Psi(0)\rangle$ is the quantum state at time $t=0$. The time evolution on a quantum circuit is performed by implementing the Trotterization protocol on $ibm\_sherbrooke$ for a small system with a length of \(L=10\). Due to the noisy qubits and imperfect gate operations, which introduce uncontrolled noise, we limit our study to short-time dynamics. To achieve a shallower depth circuit, we utilize the circuit recompilation method. To further improve the results, we utilize post-selection (PS) and zero-noise extrapolation (ZNE) error mitigation methods (see supplementary materials for details). We compare the results with the data obtained by exact diagonalization (ED) for a system of equivalent length. To obtain a complete picture of the system, we also perform ED calculations for systems up to $L=50$ sites.

{\em Results.-}
In this section, we discuss our main finding. 
For the QW, we prepare an initial state by flipping two NN spins on a ferromagnetic chain of $\downarrow$ spins, which is given as 
\begin{equation}
    |\Psi(0)\rangle = \hat{\sigma}_i^+\hat{\sigma}_j^+|....\downarrow\downarrow\downarrow\downarrow\downarrow....\rangle.
    \label{eq:bulk_ini}
\end{equation}
where $\hat{\sigma}_i^+ = (\hat{\sigma}_i^x+i\hat{\sigma}_i^y)/2$ is the spin flip operator and $i=L/2-1$ and $j=L/2$. 
\begin{figure}[t!]
    \centering
\includegraphics[width=1\columnwidth]{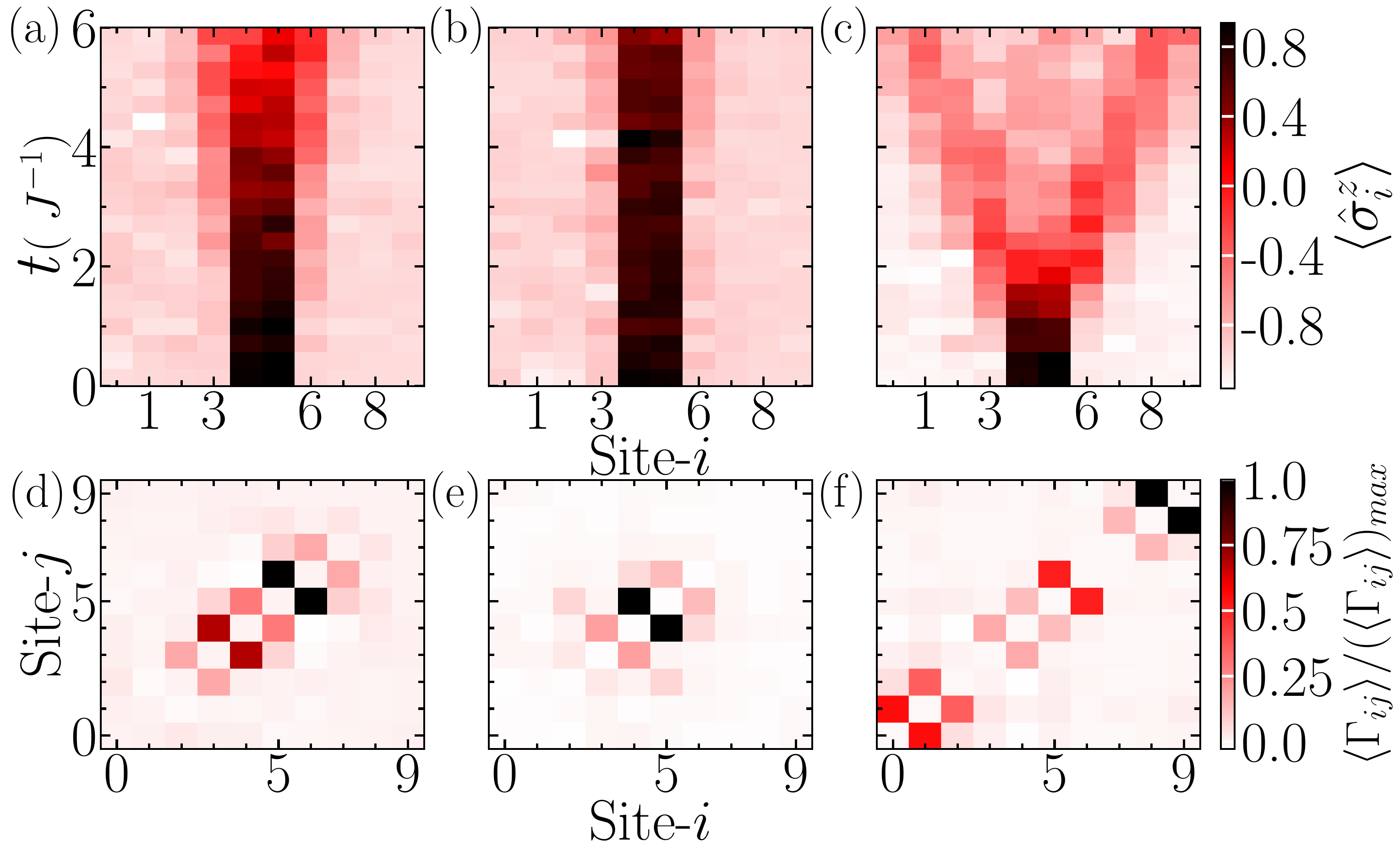}
    \caption{(a-c) The quantum computing results of the time evolved expectation value of $\hat{\sigma}^z_i$ for $J_2 = 0,~ 0.1,$ and $0.4$. (d-f) density-density correlation matrix $\langle \Gamma_{ij}\rangle$ is plotted at $t=6(~J^{-1})$ for same $J_2$ values considered in (a-c). In all the cases, we fix $J_z = 10$ and the data is PS and ZNE mitigated.}
    \label{fig:bulk_den_evo}
\end{figure}

We investigate the dynamics of such an initial state in the strong $J_z$ regime by considering $J_z=10$, for which the two NN $\uparrow$ spins are expected to from a NN bound pair ($\uparrow\uparrow$) - also known as the magnon bound state~\cite{Bloch2013, Florian2023, CorinnaKollath2024, Tonegawa1970}. In the absence of the NNN coupling ($J_2=0$), it has been shown that the magnon bound pair exhibits a dynamics with a reduced coupling strength $J_{eff} = \frac{J_1^2}{J_z}$. The spreading up of the two-magnon state is shown as the time evolution of the z-component of the spins ($\langle\sigma_i^z\rangle$) in Fig.~\ref{fig:bulk_den_evo}(a) for $J_2=0$. To confirm the formation of a magnon-bound state, we calculate the density-density correlation, defined as $\Gamma_{ij} = \langle \hat{n}_i\hat{n}_j \rangle -  \langle \hat{n}_i \rangle \delta_{ij}$
where $\hat{n}_i = \frac{1}{2}(1+\hat{\sigma}_i^z)$. The plot of $\Gamma_{i,j}$ at $t=6J_1^{-1}$ in the QW is shown in Fig.~\ref{fig:bulk_den_evo}(d) for the same parameter values as in Fig.~\ref{fig:bulk_den_evo}(a). The only finite values of the correlation matrix elements $\Gamma_{i,i+1}$ and $\Gamma_{i,i-1}$ in Fig.~\ref{fig:bulk_den_evo}(d) clearly indicate the dynamics of the NN magnon bound state.

The scenario becomes interesting with the onset of the NNN coupling $J_2$. Surprisingly, we observe that with gradual increase in $J_2$, the spreading of the two-magnon wavefunction narrows down and reaches a minimum near $J_2\sim 0.1$, which is shown in Fig.~\ref{fig:bulk_den_evo}(b). Near this point, the dynamics resembles the QW of almost localized states. 
%This indicates a slower dynamics of the two-magnon state for finite values of $J_2$ as compared to the case of $J_2=0$. 
However, with further increase in $J_2$, a faster spreading reappears, which can be seen for $J_2=0.4$ in Fig.~\ref{fig:bulk_den_evo}(c). Such a slowing down of the particle dynamics completely contradicts our conventional understanding, where further neighbor transverse coupling in the lattice is expected to reveal faster spreading of the wavefunction. Note that for both the cases of finite $J_2$ shown in Fig.~\ref{fig:bulk_den_evo} (b) and (c), the two magnons remain in the form of NN bound state, which can be seen from the correlation matrix plotted in Fig.~\ref{fig:bulk_den_evo}(e) and (f) for $J_2=0.1$ and $J_2=0.4$, respectively.  
%This can be seen from the, As soon as the we start to increase NNN XX coupling from zero,  the spreading of particles slows down and the critical slowdown in maximum when $J_2=0.1$ (see Fig.~\ref{fig:bulk_classical}(b)). With further increments, we can observe faster spreading again from Fig.~\ref{fig:bulk_classical}(c) which is plotted for $J_2=0.4$.
In the following, we will quantify this finding in detail and provide arguments supporting this non-trivial slowing down. 

\begin{figure}[t]
    \centering
    \includegraphics[width=1.0\columnwidth]{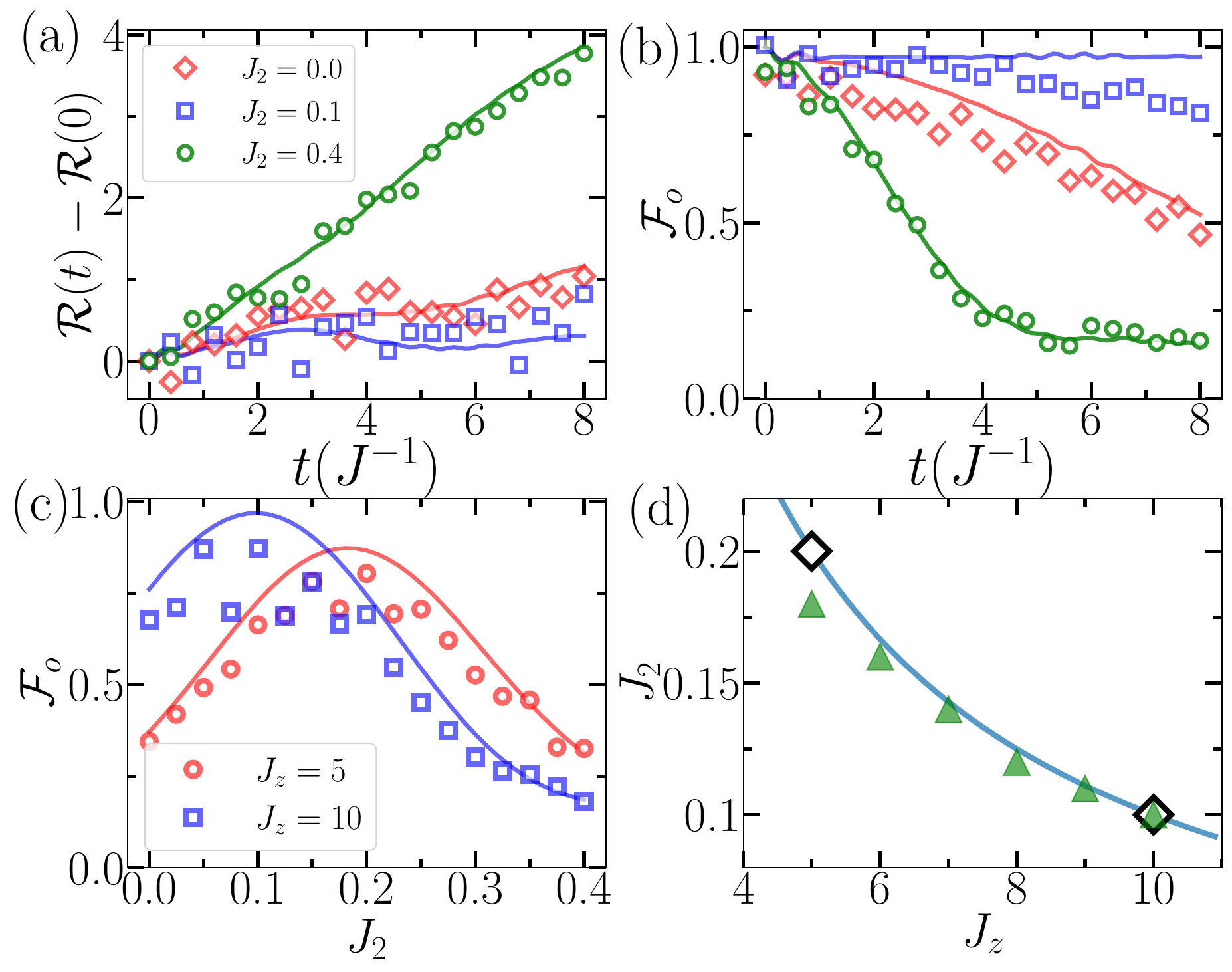}
    \caption{(a) and (b) Show the RMSD ($\mathcal{R}$) and occupation fidelity ($\mathcal{F}_o$) as a function of $t(J^{-1})$ for $J_2=0,~0.1,$ and $0.4$. (c) $\mathcal{F}_o$ at $t=5(J_1^{-1})$ is plotted as a function of $J_2$ for $J_z=5$ (red circles) and $J_z=10$ (blue squares). (d) $J_2$ corresponding to maxima of $\mathcal{F}_o$ at $t=5(J^{-1})$ for different values of $J_z$. The green triangles show the ED data, hollow markers are data from quantum computation. The solid line shows the relation $J_2=1/J_z$, which is obtained from the perturbative calculation (see text for details). For all the cases we consider $L=10$ and the initial state given in Eq.~\ref{eq:bulk_ini}.  The hollow markers are the PS and ZNE mitigated quantum computing data, and the solid lines are the from the ED calculations.}
    \label{fig:FO}
\end{figure}
To quantify the above non-monotonous behavior in the QW, we first compute the root mean square displacement (RMSD), defined as
$\mathcal{R}(t) = \sqrt{\sum_{l=0}^{L-1}(l-l_0)^2\langle \hat{n}_l(t) \rangle}$.
with $l_0=(L-1)/2$.  
In Fig.~\ref{fig:FO}(a), we plot $\mathcal{R}(t) - \mathcal{R}(0)$ as a function of time for different values of $J_2$. Note that we subtract $\mathcal{R}(0)$ from $\mathcal{R}(t)$ to discard the spatial extent of the initial state at $t=0$. For $J_2=0$ (red diamond), a linear but slower increase in RMSD is observed. However, for $J_2=0.1$ (blue square), the increase in RMSD with time is almost suppressed, indicating very slow dynamics. For $J_2=0.4$ (green circles), however, the spreading becomes faster again, indicating the revival in the dynamics.
To compare the signatures obtained from the quantum computing simulations, we plot the ED data (solid curves) for an equivalent system. The ED data convincingly confirm our findings.

To estimate the robustness of this slow dynamics, we compute the occupation fidelity~\cite{koh2022},
    $\mathcal{F}_o =\frac{1}{N} |\bold{O}_0^T \bold{O}_t|$,
where, $N$ is the number of spin excitations, $\bold{O}_0^T = [\langle \hat{n}_0\rangle_0 \langle \hat{n}_1\rangle_0......\langle \hat{n}_{L-1}\rangle_0]$ and $\bold{O}_t^T = [\langle \hat{n}_0\rangle_t \langle \hat{n}_1\rangle_t......\langle \hat{n}_{L-1}\rangle_t]$. We define the quantity in such a way that $\mathcal{F}_o\sim 1$ indicates maximum overlap with the initial state, i.e., no spreading of the spin excitations. Fig.~\ref{fig:FO}(b) shows the quantum computing data for $\mathcal{F}_o$ as a function of $t(J_1^{-1})$ for different values NNN coupling~($J_2$). We obtain that for $J_2=0.1$ (blue squares), the occupation fidelity $\mathcal{F}_o$ remains close to one for longer time compared to that for $J_2=0.0$ ( red diamonds) and $J_2=0.4$ (green circles), indicating very suppressed dynamics of the magnon bound state ($\uparrow\uparrow$). The solid curves in  Fig.~\ref{fig:FO}(b) shows the ED data for comparison. Note that the extreme slowdown of the dynamics is clearly seen for $J_2=0.1$ from the ED data although such behavior is not fully captured from the digital quantum simulation. The deviation of quantum simulation data from the exact result is due to the Trotterization error as well as the noisy qubits and imperfect gate
operations. 
\begin{figure}[t]
    \centering
    \includegraphics[width=1\columnwidth]{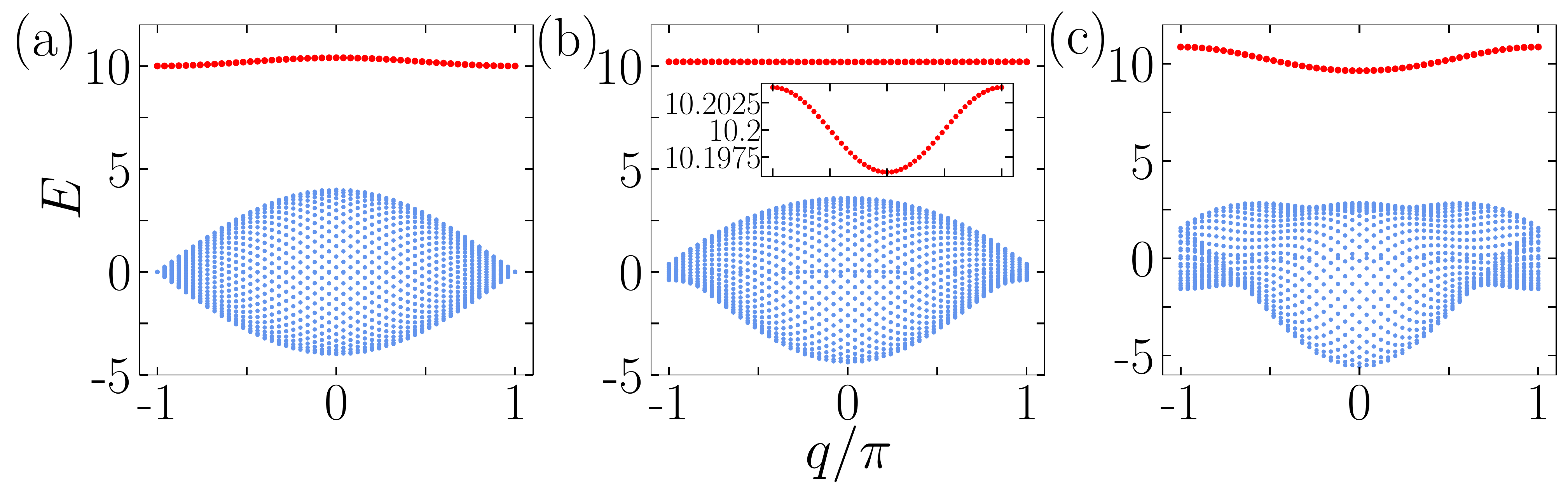}
    \caption{(a-c) The band structure of two spin excitations are shown for $J_2 = 0,~0.1$, and $0.4$, respectively. (Inset) the isolated band in (b) is zoomed-in for clarity. For all the calculations, we fix $L=50$, $J_1=1$, and $J_z=10$.}
    \label{fig:band}
\end{figure}

\begin{figure}[t!]
    \centering
    \includegraphics[width=1\columnwidth]{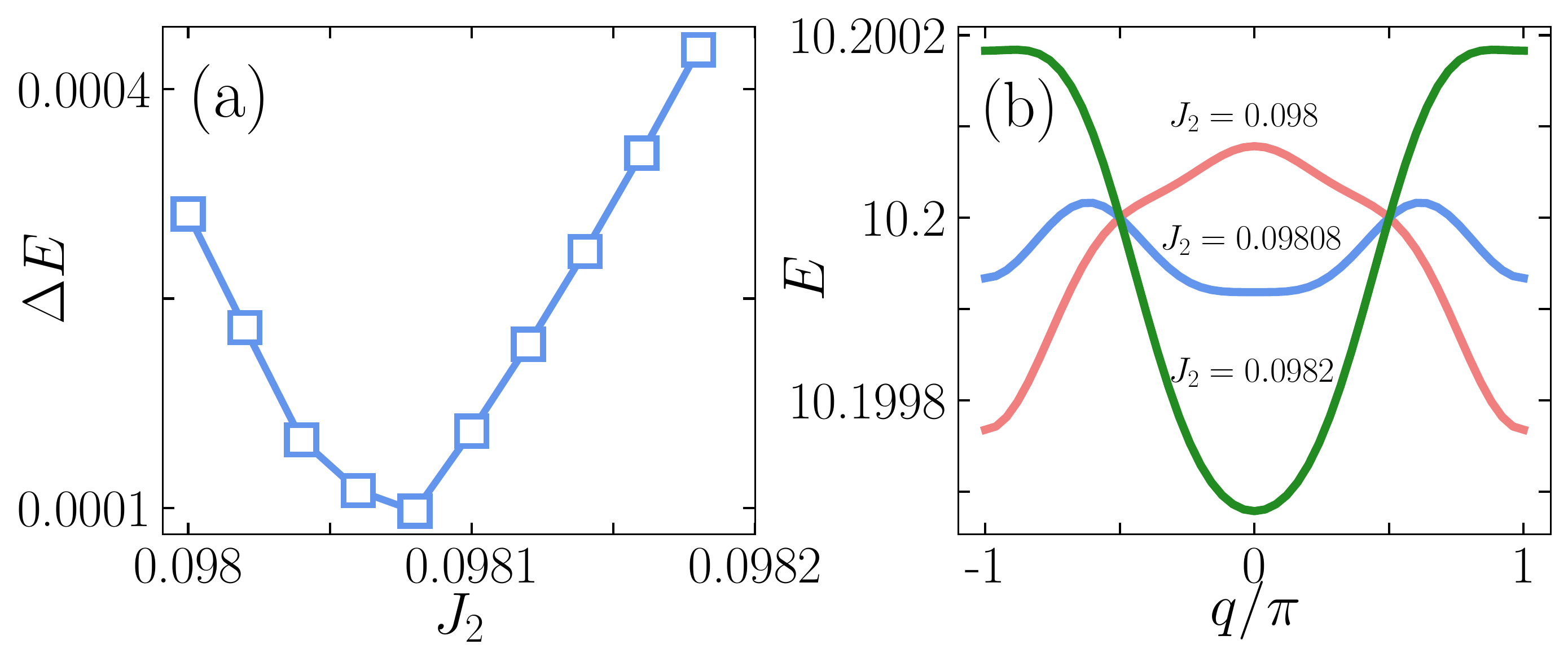}
    \caption{(a) The band width ($\Delta E$) of the isolated band is plotted as a function of $J_2$. (b) Shows the isolated bands for different values of $J_2$. All the calculations are done using ED simulations with fixed system size of $L=50$, $J_1=1$, and $J_z=10$.}
    \label{fig:band_width}
\end{figure}

To understand this behavior in the dynamics, we analyze the band structure of the system with two spin-excitations for  $J_2=0.0$, $0.1$ and $0.4$ in Fig.~\ref{fig:band} (a-c), respectively, while fixing $J_z=10$. In all three cases, typical bound state bands (red symbols) appear above the band of scattering states. The dynamics of the bound state is dictated by the dispersion of these bound state bands. Interestingly, we obtain that the width of the bound state band is narrower for $J_2=0.1$ (see inset of Fig.~\ref{fig:band}(b)) as compared to those for $J_2=0$ (Fig.~\ref{fig:band}(a)) and $0.4$ (Fig.~\ref{fig:band}(c)). The shrinking of the band for $J_2=0.1$ makes the band less dispersive and hence favors a slower dynamics of the bound state. It can also be seen that the curvature of the band flips with increase in $J_2$. To examine the nature of the band further, we plot the width of the isolated bands $\Delta E$ as a function of $J_2$ in Fig.~\ref{fig:band_width}(a) in the region near $J_2=0.1$. This clearly suggests that the band width first decreases and reaches a minimum for $J_2\sim 0.09808$, then increases again, indicating the critical $J_2$ at which the QW exhibits the slowest dynamics. Surprisingly, a closer view of the isolated bands in Fig.~\ref{fig:band_width}(b) reveal a quasi-flatband dispersion for $J_2=0.09808$ (blue curve). For comparison, we also show the bands for $J_2=0.098$ (red curve) and $J_2=0.0982$ (green curve) revealing fully dispersive bands. We infer that the emergence quasi-flatband dispersion of the bound state is responsible for the slowest dynamics of the bound state, which is reflected from the Fig.~\ref{fig:band_width} (b) as we approach $J_2=0.09808$. This also constitutes a scenario of 
disorder free quasi-localization ~\cite{Hridis2020, Park2024}. This establishes the abnormal slowing of the dynamics of the magnon bound state due to the NNN coupling. We also find that this non-trivial phenomenon occurs is in the dynamics of more than two particle bound states (see supplementary materials).

The abnormal slowing down and revival in the dynamics discussed above can be better understood from the perturbative argument by assuming the $J_z$ term as the unperturbed part and the rest of the terms in Eq.~\ref{eq:ham} as the perturbation. Using the second order perturbation theory (see supplementary materials), the effective Hamiltonian turns out to be
%{\em Analytic Calculation.-}
%We usually expect the long-range coupling to favor the spreading of the spin excitations. As we observed in the previous section, for magnon bound states, the behavior is not monotonous, and the dynamics critically slow down with NNN ZZ-coupling before the dynamics start to be faster again with the coupling strength. Here, we give an analytical argument using perturbation theory for the critical slowdown. We consider the $XX$ part as a perturbative term and the remaining part as an unperturbed part. Using second-order perturbative calculation, the effective Hamiltonian is given by,

\begin{align}
        &\hat{H}_{eff}^{(2)} = \left(-J_2+\frac{J_1^2}{J_z}\right)\sum_{i=0}^{L-2}(\hat{B}_i^\dagger\hat{B}_{i+1}+\text{H.c.}) \nonumber\\
        &+ \frac{J_2^2}{J_z}\sum_{i=0}^{L-3}(\hat{B}_i^\dagger\hat{B}_{i+2}+\text{H.c.})+\left(\frac{2J_1^2}{J_z}+\frac{2J_2^2}{J_z}\right)\sum_{i=0}^{L-2}\hat{B}_i^\dagger\hat{B}_{i},  
    \label{eq:eff_ham}
\end{align}
where $B_i=\hat{\sigma}_i^-\hat{\sigma}_{i+1}^-$. From the above Hamiltonian, it can be understood that the bound state dynamics is governed by an effective coupling of strength $-J_2+\frac{J_1^2}{J_z}$ in the leading order of $J_1$ and $J_2$. Now, if the parameters are tuned in such a way that $-J_2+\frac{J_1^2}{J_z}\sim 0$, a nearly suppressed dynamics can be observed for the bound state and this agrees well with our numerical data for $J_1=1$, $J_2=0.1$ and $J_z=10$. The expression for the effective hopping also suggests that the critical $J_2$ at which the dynamics is slowest depends as $1/J_z$. To confirm this behaviour, we plot $\mathcal{F}_o$ at $t = 5(J^{-1})$ as a function of $J_2$ for two different values of $J_z$ such as $J_z=5$ (red circles) and $J_z=10$ (blue squares) from the quantum computing simulations (see Fig.~\ref{fig:FO} (c)). The solid curves denote the results obtained from the ED calculations. The shift in the peak position or the critical $J_2$ towards the lower value with increase in $J_z$ suggests that $J_2\propto 1/J_z$. To clearly see this dependence, we plot the peak positions for different values of $J_z$ obtained from the ED simulations in Fig.~\ref{fig:FO}(d) (green triangles). The function $J_2=1/J_z$ matches well with the ED data in the perturbative limit (i.e. for higher $J_z$ limit). We also show the data from quantum computing simulations ( black squares) in Fig.~\ref{fig:FO} (d) which shows expected behaviour. 

%in the curve indicates that for large $J_z$, theIt can be seen that for both the values of $J_z$, the peak position The , in Fig.~\ref{fig:FO}(c).minimum spreading c  %imulations shown above. served from our quantum and classical computing results. In the two magnon band structure, we see that the band consisting of bound states becomes very flat for such parameter choice. This also explains that if we tune away from this parameter choice, the system exhibits re-entrant behavior in the velocity of the bound states.

% The explanation of edge localization comes from a closer look into the first term of the $\hat{H}_{eff}$. The operator involve in the first term of $\hat{H}_{eff}$ can be explicitly written as,
% \begin{equation}
%         \sum_{i=0}^{L-2}\hat{B}_i^\dagger\hat{B}_{i} = \frac{1}{4}\sum_{i=0}^{L-2}\hat{\sigma}_{i}^z\hat{\sigma}_{i+1}^z+\frac{1}{2}\sum_{i=1}^{L-2}\hat{\sigma}_{i}^z+\frac{1}{4}(\hat{\sigma}_{0}^z+\hat{\sigma}_{L-1}^z)
% \end{equation}
% The $\hat{\sigma}^{z}_i$ term acts as an onsite potential energy. 

From the effective Hamiltonian in Eq.~\ref{eq:eff_ham}, we draw an interesting inference as follows. When the bound pair is in the bulk of the chain, the effective onsite potential energy is given as $\left(\frac{2J_1^2}{J_z}+\frac{2J_2^2}{J_z}\right)$ (the last term in Eq.~\ref{eq:eff_ham}). However, if a similar calculation is performed for a system with open boundary condition with the bound state at one of the edges of the chain, then the effective onsite potential energy becomes $\left(\frac{J_1^2}{J_z}+\frac{J_2^2}{J_z}\right)$. This is exactly half of the onsite energy of the particle residing in the bulk. 
%\sout{Hence, if the QW is performed with an initial state with two spin exciations at the two NN sites at one of the edges, then they remain localized at the edge for $J_2\sim\frac{J_1^2}{J_z}$ due to the difference in the onsite potential energies between the NN sites at the edge and in the bulk.}
This difference in energy can not be overcome by the kinetic energy of the bound pair when $J_2\sim \frac{J_1^2}{J_z}$. As a consequence, we expect an edge localization in this limit.     

\begin{figure}[t!]
    \centering
    \includegraphics[width=1\columnwidth]{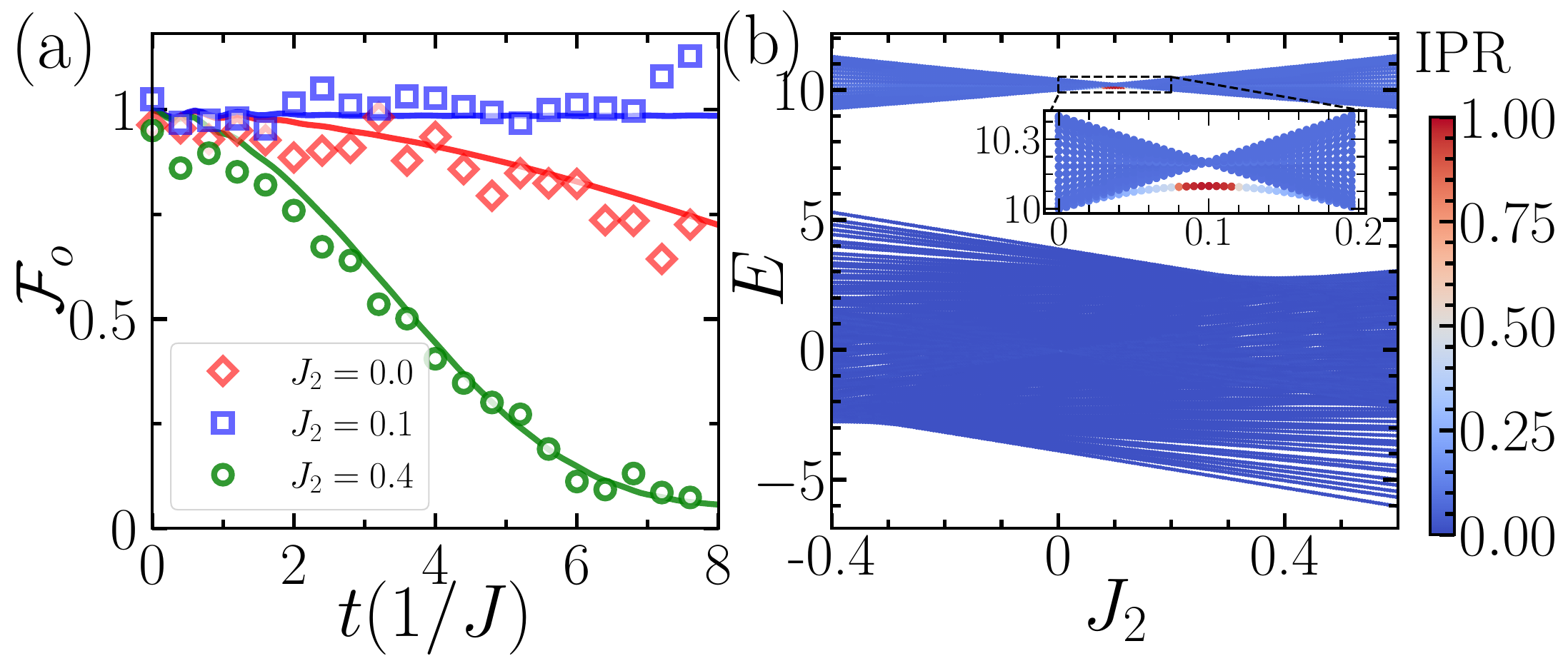}
    \caption{(a) Quantum computing results for short-time evolution of $\mathcal{F}_o$ (open symbols) after PS and ZNE mitigation are plotted for different values of $J_2$. We also plot the ED data (solid curves) for equivalent system sizes for comparison. In both the cases, we study the time evolution by exciting two magnons at the NN sites at one of the edges of the chain. (b) IPR as a function of eigenenergies and $J_2$ for a system of size $L=50$. In all the calculations we consider $J_z=10$ and $J_1=1$.}
    \label{fig:edge_overlap+Espec}
\end{figure}
%{\em Edge localization.-} 
%Often it was observed the edge of a one-dimensional system behaves quite differently from the bulk under open boundary conditions. 
To confirm this effect, we study the dynamics of the system by considering an initial state 
\begin{equation}
    |\Psi(0)\rangle = \hat{\sigma}_L^+\hat{\sigma}_{L-1}^+|....\downarrow\downarrow\downarrow\downarrow\downarrow\rangle = |....\downarrow\downarrow\downarrow\downarrow\uparrow\uparrow\rangle.
    \label{eq:edge_ini}
\end{equation}
where the two edge spins are flipped. 
%In this case, also we study the dynamics in the limits of strong $J_z$. Similar to the bulk case here we observe $\uparrow\uparrow$ bound pair dynamics in strong $J_z$ limit. In Fig.~\ref{fig:edge_overlap+Espec}~(a) we plot the classical simulation results of occupation fidelity as a function of time when we start from the initial state given in Eq.~\ref{eq:edge_ini} for system size $L=50$. 
We plot the time evolution of the occupation fidelity $\mathcal{F}_o$  in Fig.~\ref{fig:edge_overlap+Espec}(a)  for $J_z=10$ and for different values of $J_2$. We obtain that $\mathcal{F}_o$ remains close to one as a function of $t(J^{-1})$ for $J_2=0.1$ (blue squares). This behaviour can also be seen from the ED data (solid curve) with which the quantum computing data qualitatively matches well. This indicates a complete suppression of the dynamics of the magnon bound state leading to a scenario of edge localization. For other values of $J_2=0.0~(< 0.1)$ (red diamonds) and $J_2=0.4~(> 0.1)$ (green circles) the magnon bound state spreads into the bulk which is indicated by the decay of $\mathcal{F}_o$ as a function of $t(J^{-1})$. This observation is also consistent with the evolution of the spin-excitation (see supplementary materials). The edge localization of the magnon bound state is further confirmed from the inverse participation ratio, $\text{IPR} = \sum_i |\psi_i|^4$, which measures the extent of localization of the states. A non-zero IPR indicates a localized state and for a fully localized state, the IPR $\sim 1$~\cite{hobbyhorse}. We plot the $\text{IPR}$ of all the states obtained by diagonalizing the Hamiltonian in Eq.~\ref{eq:ham} as a function of their eigenenergies and $J_2$ in Fig.~\ref{fig:edge_overlap+Espec}(b) for $J_z=10$. The edge localization is confirmed by the $IPR \sim 1$ of the isolated eigenstate states (see inset) for $J_2 \sim 0.1$.

{\em Conclusion.-}
We studied the dynamics of a bound state in a one dimensional lattice with both NN and NNN hopping. By mapping the system to a spin-1/2 chain with NNN transverse coupling, we show that for appropriate choice of the coupling strengths the bound state exhibits an abnormally slow dynamics when the bound state is inside the bulk of the lattice. We demonstrate that while the dynamics of a single particle in such a scenario is expected to be faster, the counter intuitive behavior in the case of bound state is due the onset of a quasi-flat band dispersion. Additionally, we obtained that when the bound state is in the edge of an open chain, the dynamics is fully suppressed. We analyzed the dynamics using a NISQ device and complement the results through ED calculation and also using the perturbative approach. 

Our finding reveals a non-trivial and counter-intuitive behavior in the dynamics of interacting particles by going beyond our existing understanding. This also opens up directions for further exploration in the context of other lattice models and in the limit of true many-body systems. Given the recent interest in the topic including experiments involving few particle dynamics and the simplicity of the model, our findings can in-principle be observed in an NNN coupled systems such as ultracold atoms in optical lattices~\cite{Bloch2013} or superconducting circuits~\cite{Google2022} or trapped ion arrays~\cite{Knap2023}. 

{\em Acknowledgement.-}
We thank Immanuel Bloch, S. D. Mahanti and Diptiman Sen for useful discussions. T.M. acknowledges support from the Science and Engineering Research Board (SERB), Govt. of India, through project No. MTR/2022/000382 and STR/2022/000023.

\bibliography{references}

\clearpage
\onecolumngrid
\begin{center}
\textbf{Supplementary materials for 
 ``Anomalous slow-down of the bound state dynamics in a non-locally coupled quantum
circuit"}
\end{center}

    In this supplementary material, we provide a detailed description of the circuit implementation of our model, circuit optimization, and error mitigation techniques. We also include some additional results to clearly visualize the edge localization. Furthermore, we extend our numerical analysis to a three-particle bound state and support our results with a perturbative argument.  
    Finally, we provide a brief discussion of the perturbative method.

\vspace{5mm}

\twocolumngrid

\section{Circuit Implementation}
In this section, we briefly discuss the quantum circuit implementation of the time evolution operator.
\subsection{Time evolution on Quantum Circuit}
A suitable method for implementing the time-evolution operator in quantum circuits is the Suzuki-Trotter decomposition~\cite{Sieberer2019, Heyl2019, Smith2019}. The first-order Suzuki-Trotter decomposition of the time evolution operator is given by,
\begin{align}
    \hat{U}(t) &= e^{-i\hat{H}t}= \Big(e^{-i\hat{H}\Delta t}\Big)^n \nonumber \\
    &= \Big(e^{-i\hat{H}_0\Delta t}e^{-i\hat{H}_1\Delta t}e^{-i\hat{H}_2\Delta t}\Big)^n + \mathcal{O}(n\Delta t^2),
\end{align}
where,
\begin{equation}
\begin{split}
    &\hat{H}_0 = -\frac{J_1}{2}\sum_{j=0}^{L-2} (\hat{\sigma}_{j}^x \hat{\sigma}_{j+1}^x+\hat{\sigma}_{j}^y \hat{\sigma}_{j+1}^y)+\frac{J_z}{4}\sum_{j=0}^{L-2} \hat{\sigma}_j^z\hat{\sigma}_{j+1}^z,\\
    &\hat{H}_1 = -\frac{J_2}{2}\sum_{j=0}^{L-3} (\hat{\sigma}_{j}^x \hat{\sigma}_{j+2}^x+\hat{\sigma}_{j}^y \hat{\sigma}_{j+2}^y),\\
    & \hat{H}_2 = \frac{J_z}{4} \sum_{j=0}^{L-2} (\hat{\sigma}_j^z+\hat{\sigma}_{j+1}^z)
\end{split}
\end{equation}
and $n$ is the number of the Trotter steps with $\Delta t = t/n$. We rewrite $\hat{U}(t)$ for $n$ number of Trotter steps as,
\begin{align*}
    \hat{U}(t) &= \Big(\prod_{j=even}\hat{A}_{j}\prod_{i=odd}\hat{A}_{j}\prod_{j=0}^{L-3}\hat{B}_{j}\prod_{j=0}^{L-1}\hat{C}_{j}\Big)^{n}+\mathcal{O}(n\Delta t^2).
\end{align*}
To make the notation simpler, we use,
\begin{equation}
\begin{split}
    \hat{A}_j&=e^{-i\Delta t\left[-J_1/2 (\hat{\sigma}_{j}^x \hat{\sigma}_{j+1}^x+\hat{\sigma}_{j}^y \hat{\sigma}_{j+1}^y)+J_z/4( \hat{\sigma}_j^z\hat{\sigma}_{j+1}^z)\right]},\\
    \hat{B}_j&=e^{-i\Delta t\left[-J_2/2 (\hat{\sigma}_{j}^x \hat{\sigma}_{j+2}^x+\hat{\sigma}_{j}^y \hat{\sigma}_{j+2}^y)\right]},\\
    \hat{C}_j &= e^{-i\Delta t\left[J_z/4  (\hat{\sigma}_j^z+\hat{\sigma}_{j+1}^z)\right]}.
\end{split}
\end{equation}
The first-order Trotter circuit for a single Trotter step is shown in Fig.~\ref{fig:ckt}(a). Fig.~\ref{fig:ckt}(c) and (d) show the quantum circuit for the expression $\hat{A}_j$ and $\hat{B}_j$ with the optimal number of gates~\cite{vatan2004,Smith2019} and the $\hat{C}_j$ is just single qubit $R_z(J_z\Delta t/2)$ rotational gate. The error depends on the time step $\Delta t$, which can be reduced by making the step size smaller or using higher-order Trotter decomposition. However, this will increase the number of noisy quantum gates in the circuit, which will severely affect the quality of the results. To avoid this, we consider a moderately small $\Delta t=0.1$ and first-order Suzuki-Trotter decomposition in our calculation. 

\subsection{Circuit recompilation}
In the Trotterization method discussed above, the circuit depth increases linearly with number of Trotter steps. This results in an increase in the number of noisy quantum gates, which affects the quality of the results. To overcome this issue, we use the circuit recompilation method to construct a constant-depth circuit~\cite{Jones2022robustquantum, Khatri2019quantumassisted, heya2018variational}.

\begin{figure*}
    \centering\includegraphics[width=2\columnwidth]{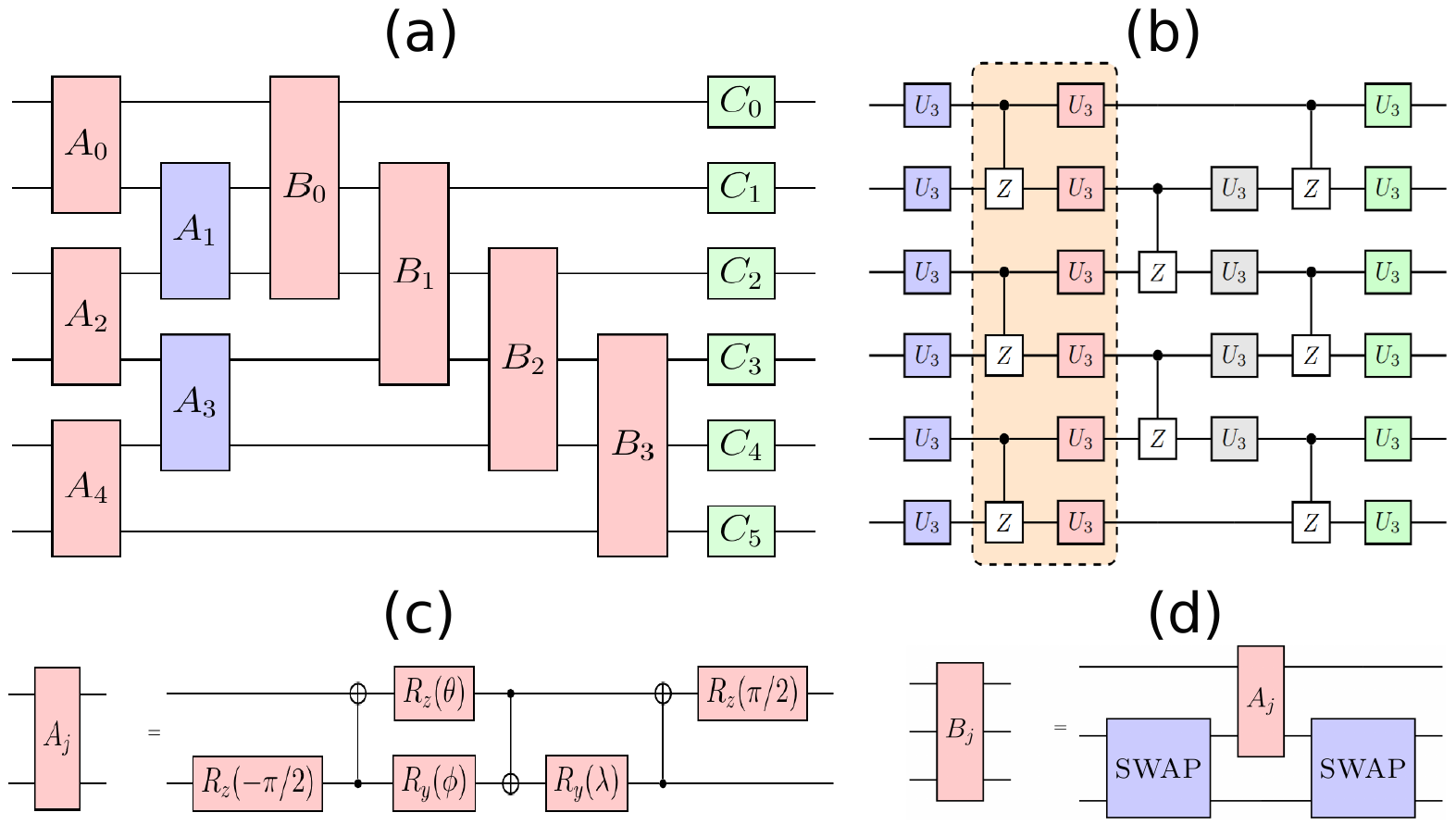}
    \caption{(a) The circuit diagram for a single Trotter step. (b) The recompiled circuit consisting of an alternative single-qubit $U_3$ layer and two-qubit control-Z layer. A gate round consists of a single layer of single-qubit gate and a single layer of two-qubit gate shown in the dashed box. (c) and (d) Explicit circuit diagram for the operators $A_{j}$ and $B_j$, respectively. In (c), we use $\theta=\frac{\pi}{2}-\frac{J_z\Delta t}{2}$, $\phi=J_1\Delta t-\frac{\pi}{2}$, $\lambda=\frac{\pi}{2}-J_1\Delta t$ and for (d) we consider $\theta=\frac{\pi}{2}$, $\phi=J_2\Delta t-\frac{\pi}{2}$,  $\lambda=\frac{\pi}{2}-J_2\Delta t$.}
    \label{fig:ckt}
\end{figure*}

Using the tensor network based optimization method we convert our Trotter circuit into a parametrized quantum circuit shown in \ref{fig:ckt} (b). We use the QUIMB library to optimize circuit parameters~\cite{quimb}. Our parametrized circuit consists of alternating layers of single-qubit $U_3$ and two-qubit control-Z (CZ) gates. Where $U_3$ contains three adjustable parameters given by,
\begin{equation}
U_3(\theta, \phi, \lambda)=
    \begin{pmatrix}
        \cos(\frac{\theta}{2}) & -e^{i\lambda}\sin(\frac{\theta}{2}) \\
        e^{i\phi}\sin(\frac{\theta}{2}) & e^{i(\phi+\lambda)}\cos(\frac{\theta}{2})
    \end{pmatrix}.
\end{equation}
We can get the optimized parametrized circuit which can mimic our Trotter circuit by maximizing the following quantity,
\begin{equation}
    F(\Theta) = \Big{|}\langle \Psi(0)|\hat{U}_{rec}(\Theta)^\dagger \hat{U}_{tar}|\Psi(0)\rangle\Big{|}^2.
\end{equation} 
We denote our Trotter circuit by $\hat{U}_{tar}$ and our parametrized circuit by $\hat{U}_{rec}$, and all the rotation parameters of $U_3$ gates are denoted by $\Theta$. For our study, we consider the parametrized circuit with eight alternative layers of $U_3$ and CZ gates, and we obtain the parametrized circuit fidelity $F(\Theta)> 99\%$. In this parametrized circuit with a relatively smaller number of gate counts, we still expect some undesirable noise in the output at this stage. To reduce this noise effect, we use two error mitigation methods discussed in the following section.

\section{Error mitigation}

% To construct the quantum circuit we use the Trotterization method. However, the problem with this method is the number of noisy quantum gates increases with time steps. We use the circuit recompilation(a brief discussion given in the main text) method to reduce the circuit depth but still a sufficient number of noisy quantum gates appear in our circuit which reduces the accuracy of the results. We use the following error mitigation method to increase the accuracy of our results.

In this section, we briefly discuss two error mitigation methods implemented to improve the noisy results obtained from the quantum processor. \\

\textit{Post-Selection Method:}
Respecting the symmetry of the system one could implement the post-selection methods to increase the accuracy of the results by discarding the undesirable states at the outputs. If $\hat{O}$ is a conserved quantity of our system, then throughout the time evolution process that quantity is conserved i.e. $\langle \hat{O} \rangle_{t=0}=\langle \hat{O} \rangle_{t=t_f}$, where $t_f$ is the final time. The Hamiltonian we consider conserves the number of spin excitations in our system which is the post-selection for our system. In this approach, the states those conserve the total number of spin excitations are considered to be valid and the rest are ignored.\\

\textit{Zero Noise Extrapolation Method:}
ZNE is a suitable error mitigation method for implementation on near-term hardware to improve the accuracy of the results by suppressing the incoherent noise. The basic idea is to extrapolate the noisy data in the zero noise limit. Two major steps are involved in this procedure $(i)$ scale-up the noise of the circuit and $(ii)$ extrapolate the noisy data to zero noise limit.

$(i)$ Scale up the circuit noise: We first use the gate-level unitary folding~\cite{ZNE_Li_2017, ZNE_Temme_2017} method using the library mitiq~\cite{mitiq}, to scale up the noise in our circuit. We increase the circuit depth in such a way that the expectation value of the observable does not change if we consider an ideal quantum simulation. However, the qubits and gates available in recent times are not perfect, and the quality of the results decreases with an increase in circuit depth. Let $\eta$ be the noise level of the original circuit and $\eta^\prime=\lambda\eta$ be the noise of the scaled circuit. For extrapolation, we run the circuit for three different noise scales $\lambda=1$ (original circuit), $\lambda=2$ (doubling the circuit depth), and $\lambda=3$ (increasing the circuit depth threefold). As the two-qubit gates are much more noisy than the single-qubit gates, we consider only two-qubit gates for noise scaling. After getting the noisy expectation values, we extrapolate it for zero noise. The extrapolation technique is given as follows.\\

$(ii)$ Extrapolation of noisy expectation values: Let, $\hat{O}$  be the observable we want to calculate. The expectation value of the observable $\hat{O}$ for each noise scale $\eta^\prime = \lambda\eta$ is denoted by $O_\lambda$ . Once $\lambda$ and $O_\lambda$ values are known, we can easily calculate the zero noise extrapolated value of the observable ($O_{\lambda=0}$). For extrapolating into the zero noise regime, we use Richardson's extrapolation~\cite{ZNE_Temme_2017, ZNE_Kandala2019} technique.\\
% Hamiltonian in second quantization notation,

% \begin{equation}
%     \begin{split}
%         \hat{H} =t_1\sum_j(\hat{a}_j^{\dagger}\hat{a}_{i+1}+h.c.) &+ t_2\sum_j(\hat{a}_j^{\dagger}(1-2\hat{n}_{j+1})\hat{a}_{i+2}+h.c.)\\
%         &+4V\sum_j \hat{n}_j\hat{n}_{j+1}
%     \end{split}
% \end{equation}

\begin{figure}[t!]
    \centering
    \includegraphics[width=1\columnwidth]{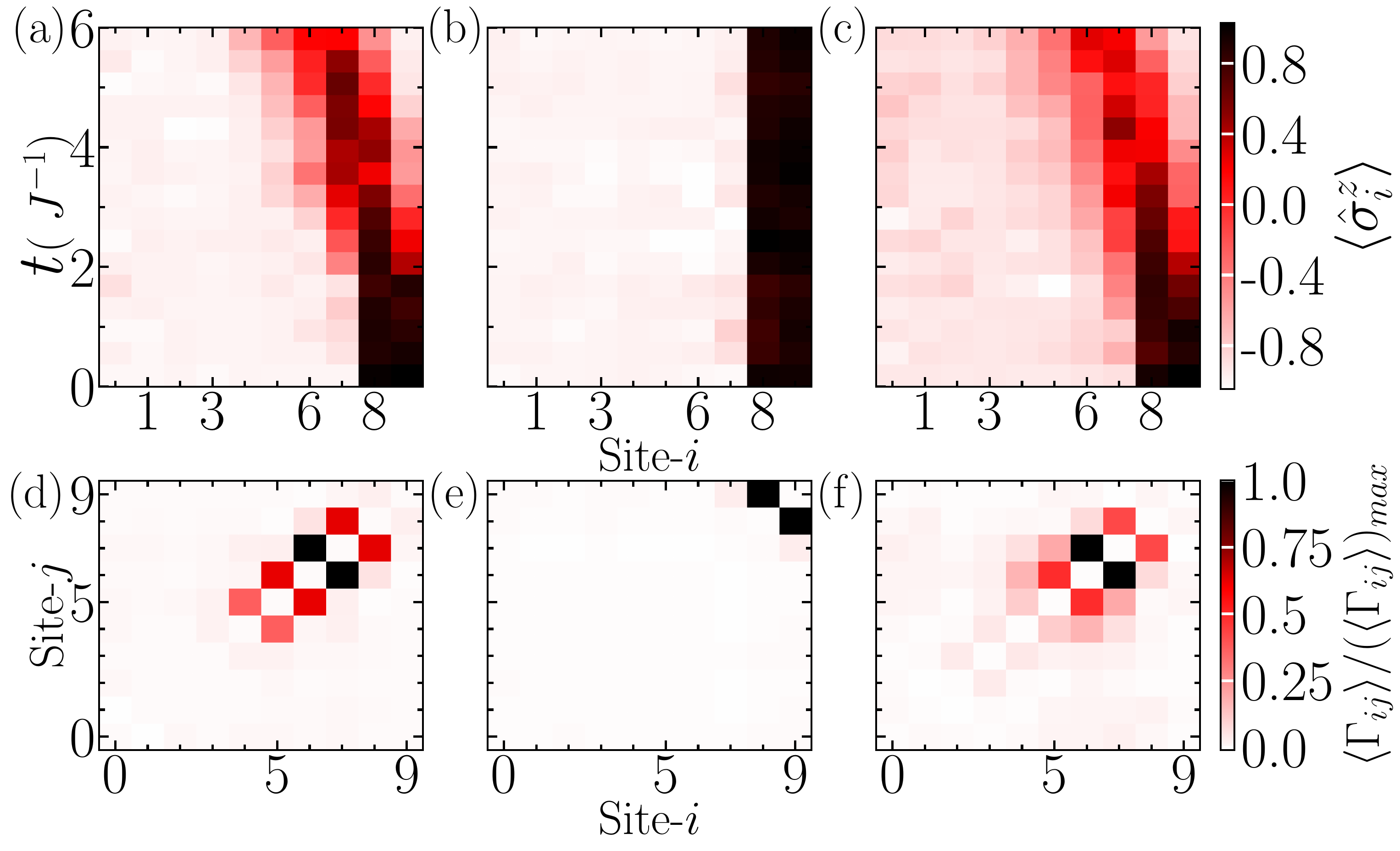}
    \caption{(a-c) The PS+ZNE mitigated data for the time evolved expectation value of $\sigma^z_i$ for $J_2 = 0,~ 0.1,$ and $0.4$. (d-f) density-density correlation function $\Gamma_{ij}$ plot at $t=6(~J^{-1})$ for same $J_2$ values considered in (a-c). We fix the other parameters of the system to $J_1=1$ and $J_z=10$.}
    \label{fig:edge_den_evo}
\end{figure}

\section{Edge localization} 
In this section, we discuss the edge localization of magnon bound state. Using insights from the quantum computing results, we demonstrate that for a specific value of $J_2$, complete edge localization can be achieved. We study the time evolution of two excitation on the two NN sites at one of the edges of the chain. We plot the time evolution of $\langle\hat{\sigma}^z_i\rangle$  in Fig.~\ref{fig:edge_den_evo} obtained from quantum computing simulations using $ibm\_sherbrooke$ for a system of size $L=10$. By considering $J_z=10$ and $J_1=1$, we observe that the bound state remain localized at the edge for $J_2=0.1$ (Fig.~\ref{fig:edge_den_evo}(b)). For other values of $J_2=0 (<0.1)$ and $J_2=0.4 (>0.1)$, the bound state spreads into the bulk which can be seen from Fig.~\ref{fig:edge_den_evo}(a) and Fig.~\ref{fig:edge_den_evo}(c) respectively. This result is also complemented by the correlation matrix $\Gamma$ displayed in Fig.~\ref{fig:edge_den_evo}(d-f).

\begin{figure}[t!]
    \centering
    \includegraphics[width=1\columnwidth]{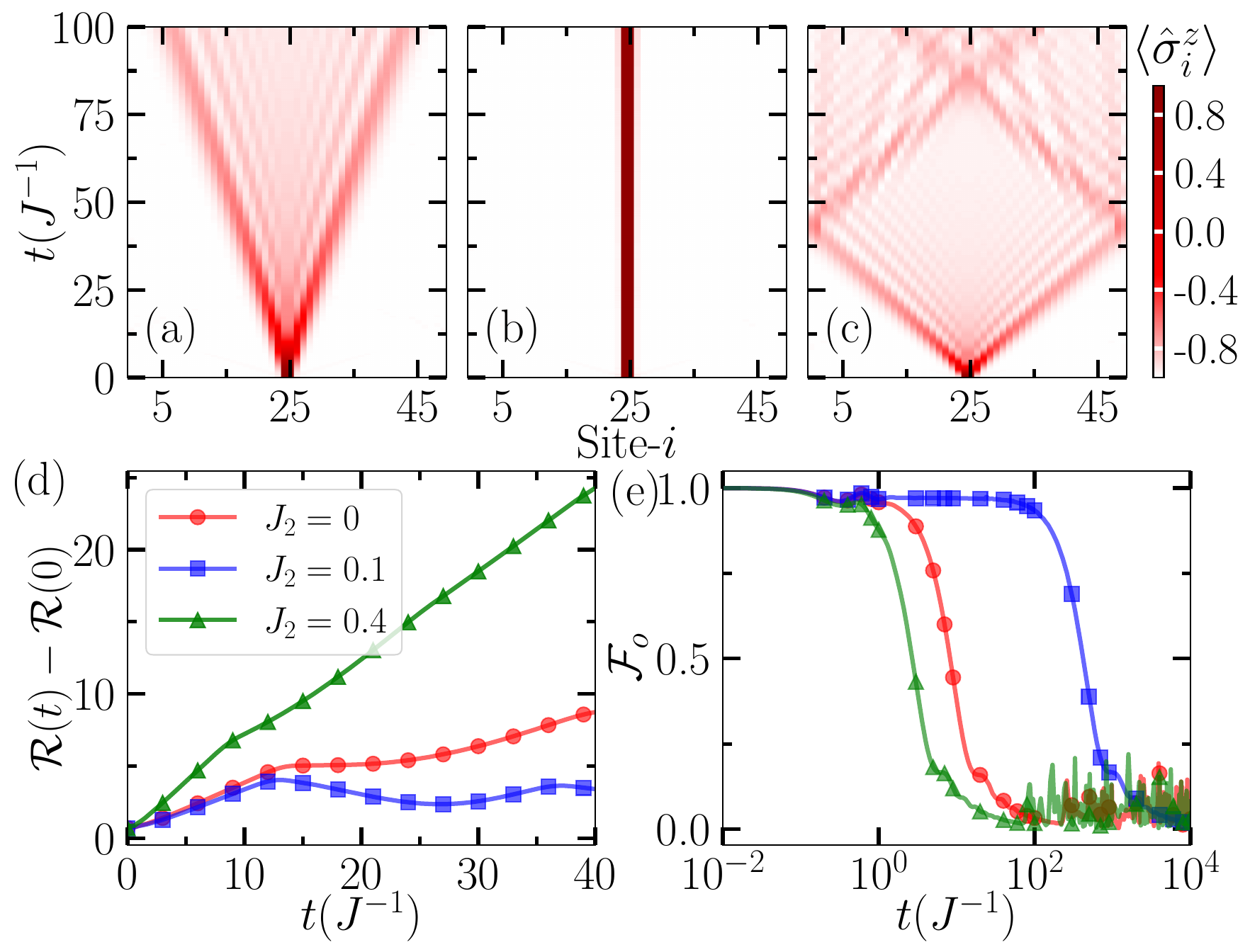}
    \caption{ ED results for a system of size $L=50$. (a), (b) and (c) show the time evolution expectation value of $\hat{\sigma}^z_i$ for $J_2=0,~0.1,$ and $0.4$ respectively. (d) $\mathcal{R}(t)-\mathcal{R}(0)$ is plotted as a function of $t(J^{-1})$ for $J_2=0,~0.1,$ and $0.4$. (e) The time evolution of $\mathcal{F}_o$ is plotted as a function of $t(J^{-1})$ for $J_2$ values as considered in (d). 
     In all cases, we study the time evolution with the initial state given in Eq.~\ref{eq:supmatini1} with  $J_z=10$ and $J_1=1$.}
    \label{fig:bulk_classical}
\end{figure}
\section{Dynamics with two and three excitations}

In this section, we present the dynamics of two and three spin excitations obtained from the ED simulations. First, we investigate the time evolution of two spin-excitations for system size $L=50$, starting from the initial state
\begin{equation}
    |\Psi(0)\rangle = \hat{\sigma}_{L/2-1}^+\hat{\sigma}_{L/2}^+|....\downarrow\downarrow\downarrow\downarrow\downarrow\downarrow....\rangle=|....\downarrow\downarrow\uparrow\uparrow\downarrow\downarrow....\rangle,
    \label{eq:supmatini1}
\end{equation}
where $\hat{\sigma}_i^+ = (\hat{\sigma}_i^x+i\hat{\sigma}_i^y)/2$ is the spin flip operator. Fig.~\ref{fig:bulk_classical}(a-c) shows the bulk dynamics for $J_2=0,~0.1,$ and $0.4$, respectively, while keeping the $J_z=10$ and $J_1=1$. For $J_2=0.1$, we observe suppressed dynamics of the bound state compared to other parameter values considered, which agrees well with our digital quantum simulation as well as the perturbative result discussed in the main text. To further quantify the spreading of excitations, we plot $\mathcal{R}(t)-\mathcal{R}(0)$ in Fig.~\ref{fig:bulk_classical}(d). The value of $\mathcal{R}(t)-\mathcal{R}(0)\sim0$ for $J_2=0.1$ indicates the suppressed spreading of spin excitation, whereas in other cases~($J_2=0,~\text{and}~0.4$) we can see a relatively faster spreading. To quantify the long-time behaviour, we plot the occupation fidelity $\mathcal{F}_o$ as a function of $t(J^{-1})$ for $J=0$ (red circles), $J=0.1$ (blue squares), and $J_2=0.4$ (green triangles) in Fig.~\ref{fig:bulk_classical}(e). The value of $\mathcal{F}_o\sim 1$ up to a long-time dynamics ($t \sim 100$) for $J_2=0.1$  and subsequent slow decay towards $\mathcal{F}_o\sim 0$ indicate the suppressed dynamics of the bound state.

\begin{figure}[t!]
    \centering
    \includegraphics[width=1\columnwidth]{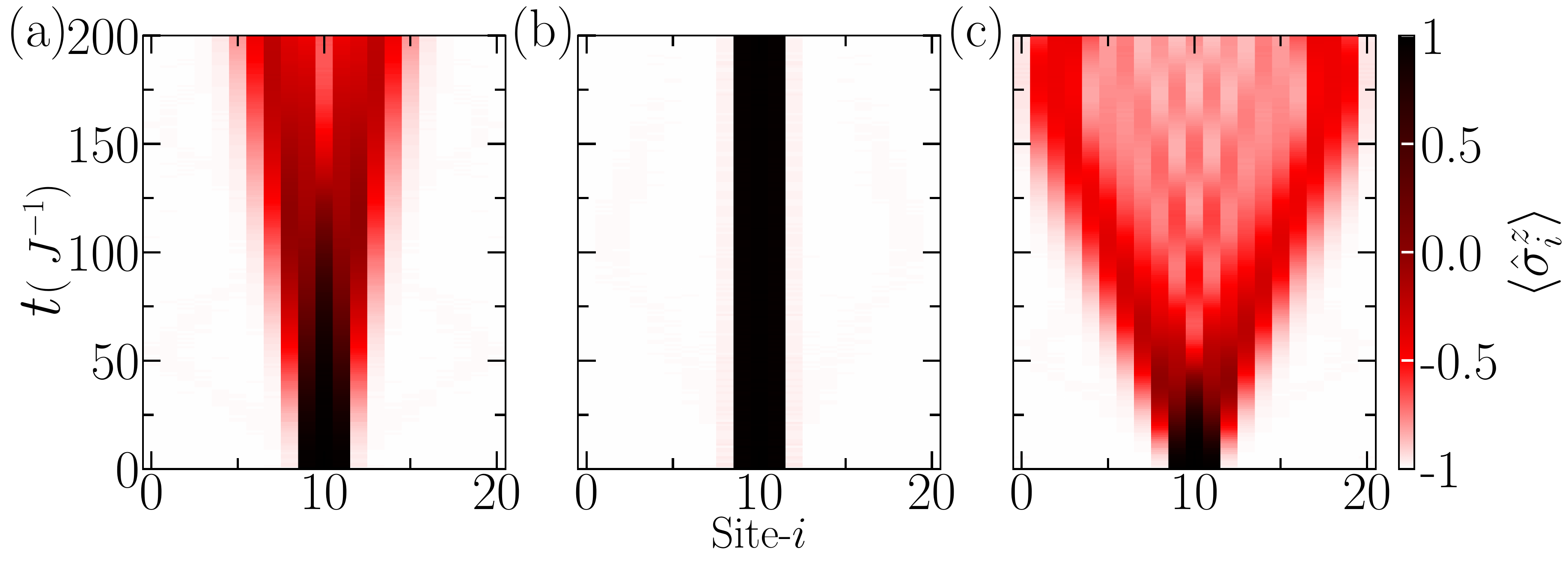}
    \caption{(a-c) the time evolved expectation value of $\hat{\sigma}^z_i$  is plotted for $J_2 = 0,~ 0.05,$ and $0.2$. We fix the other parameters of the system as $J_1=1$ and $J_z=10$.}
    \label{fig:den_evo_3_part}
\end{figure}

We also obtain similar behaviour in the dynamics of three spin excitations at the middle of the lattice. The dynamics is shown in Fig.~\ref{fig:den_evo_3_part} with the initial state,
\begin{equation}
    |\Psi(0)\rangle = \hat{\sigma}_{L/2-1}^+\hat{\sigma}_{L/2}^+\hat{\sigma}_{L/2+1}^+|...\downarrow\downarrow\downarrow\downarrow\downarrow...\rangle=|...\downarrow\uparrow\uparrow\uparrow\downarrow...\rangle,
\end{equation}
and strong $J_z$ ensuring a three particle bound state ($\uparrow\uparrow\uparrow$). We obtain that the dynamics of the three particle bound state is substantially suppressed for $J_2=0.05$ when $J_z = 10$, as shown in Fig.~\ref{fig:den_evo_3_part} (b). However, for other values of $J_2$, we obtain comparatively faster spreading (see Fig.~\ref{fig:den_evo_3_part} (a) and (c)). 
%the three particle bound state spreads in the lattice with an effective hopping strength $J_{eff} = \frac{J_1^3}{J_z^2}$. Like two excitation dynamics cases, the system shows more reacher dynamics because of the inclusion of $J_2$. Here we also can see a non-monotonous behavior of particle spreading as a function of $J_2$. From Fig.~\ref{fig:den_evo_3_part}(b) for $J_2=0.05$, we can see the suppression of particle spreading as compared to Fig.~\ref{fig:den_evo_3_part}(a) where $J_2=0$. Further, if we increase the $J_2$ value we can see the spreading of particles again, which can be seen from Fig.~\ref{fig:den_evo_3_part}(c) where we fix $J_2 = 0.2$. We can explain the slowing down of the 3-particle bound pair dynamics at a specific value of $J_2$, in a similar line to the two-particle case, using band structure (not shown) and the perturbative argument. 

Here we calculate the effective Hamiltonian by considering all the $J_1$ and $J_2$ coupling as the perturbative terms, and the rest of the terms as the unperturbed part. Using the third-order perturbative calculation, the effective Hamiltonian is given by,
\begin{eqnarray}
        \hat{H}_{eff}^{(3)} &=& \frac{-J_1}{J_z}\left(-2J_2+\frac{J_1^2}{J_z}+\frac{4J_2^2}{J_z}\right)\sum_{i=1}^{L-3}(\hat{B}_i^\dagger\hat{B}_{i+1}+\text{H.c.}) \nonumber\\
        &+& \frac{J_2^3}{J_z^2}\sum_{i=1}^{L-4}(\hat{B}_i^\dagger\hat{B}_{i+2}+\text{H.c.})\nonumber\\
        &+& \left(\frac{2J_1^2}{J_z}+\frac{4J_2^2}{J_z}-\frac{6J_1^2J_2}{J_z^2}\right)\sum_{i=1}^{L-2}\hat{B}_i^\dagger\hat{B}_{i},
        \label{eq:3part_eff_ham}
\end{eqnarray}
where $B_i=\hat{\sigma}_{i-1}^-\hat{\sigma}_i^-\hat{\sigma}_{i+1}^-$. The bound state exhibits motion with an effective coupling strength given by $-\frac{J1}{J_z}(-2J_2+\frac{J_1^2}{J_z})$ at the leading order in $J_1$ and $J_2$. If the parameters are tuned such that $-2J_2+\frac{J_1^2}{J_z}=0$, the bound state is expected to undergo a critical slowdown. This suggests that the effective Hamiltonian agrees well with our numerical results shown in Fig.~\ref{fig:den_evo_3_part}.

\section{Perturbative calculation}
In the main text, we show that in the limit $J_z\gg J_1,~J_2$, two spin excitation forms a bound pair and behave as a single composite particle and move in the lattice with a reduced effective mass. In this limit, we can use the perturbative approach following the prescription given in Refs.~\cite{Takahashi_1977,Cai2021}, to study the behavior of the composite particle. By treating $J_z$ as the unperturbed part and the rest of the terms perturbatively we can  write the Hamiltonian as,
\begin{equation}
    \hat{H} = \hat{H}_0 + \hat{H}_1,
\end{equation}
where
\begin{equation}
    \hat{H}_0 = \frac{J_z}{4}\sum_{j=0}^{L-2} (1+\hat{\sigma}_j^z)(1+\hat{\sigma}_{j+1}^z)
\end{equation}
\begin{equation}
    \hat{H}_1 = -\frac{J_{1}}{2}\sum_{\langle i, j\rangle} (\hat{\sigma}_{i}^x \hat{\sigma}_{j}^x+\hat{\sigma}_{i}^y \hat{\sigma}_{j}^y)
    -\frac{J_{2}}{2}\sum_{\langle\langle i, j\rangle\rangle} (\hat{\sigma}_{i}^x \hat{\sigma}_{j}^x+\hat{\sigma}_{i}^y \hat{\sigma}_{j}^y)
    \label{eq:pert_ham}
\end{equation}
We consider two excitations in the system for which the unperturbed Hamiltonian has two possible energies. When two excitations stay at adjacent sites, the system has $L-1$ fold degeneracy with eigenenergy $E_0 = J_z$, and we denote the corresponding eigenstates by $\{|E_0^{(j)}\rangle = |j, j+1\rangle\}$. The other possible eigenenrgy is $E_1=0$, for which the eigenstate is $\{ |E_1^{(j_1j_2)}\rangle = |j_1, j_2\rangle\}$, where $j_1\neq j_2\pm1$.

The projection operator to the bound state subspace can be written as,
\begin{equation}
    P_0 = \sum_j |E_0^{(j)}\rangle\langle E_0^{(j)}|,
    \label{eq:proj0}
\end{equation}
and the projection operator to the orthogonal subspace of $P_0$ is
\begin{equation}
    P_1 = \sum_{j_1,J_2}\frac{1}{E_0-E_1}|E_1^{(j_1j_2)}\rangle\langle E_1^{(j_1j_2)}|.
    \label{eq:proj1}
\end{equation}
The second-order effective Hamiltonian can be expressed as 
\begin{equation}
    \hat{H}_{eff}^{(2)} = E_0P_0 + P_0\hat{H}_1P_0+P_0\hat{H}_1P_1\hat{H}_1P_0.
    \label{eq:pur_exp}
\end{equation}
Substituting Eq.~\ref{eq:pert_ham}, ~\ref{eq:proj0}, and \ref{eq:proj1} into Eq.~\ref{eq:pur_exp} we get the effective Hamiltonian as
\begin{align}
        &\hat{H}_{eff}^{(2)} = \left(-J_2+\frac{J_1^2}{J_z}\right)\sum_{i=0}^{L-2}(\hat{B}_i^\dagger\hat{B}_{i+1}+\text{H.c.}) \nonumber\\
        &+ \frac{J_2^2}{J_z}\sum_{i=0}^{L-3}(\hat{B}_i^\dagger\hat{B}_{i+2}+\text{H.c.})+\left(\frac{2J_1^2}{J_z}+\frac{2J_2^2}{J_z}\right)\sum_{i=0}^{L-2}\hat{B}_i^\dagger\hat{B}_{i},   
\end{align}
where $B_i=\hat{\sigma}_i^-\hat{\sigma}_{i+1}^-$. This effective Hamiltonian agrees well with our numerical results for the dynamics with two excitations as the initial state.

Now, if we consider three excitations, then three possible energies corresponding to the unperturbed Hamiltonian are

\begin{enumerate}
    \item $E_0 = 2J_z$, possible only when three excitations reside at adjacent sites, and therefore correspond to the eigenstate $|E_0^{(j)}\rangle=|j,~j+1,~j+2\rangle$.
    \item $E_1 = J_z$, only two excitations at adjacent sites are possible. Therefore, corresponding states are $|E_1^{(j_1j_2)}\rangle=|j_1,~j_1+1,~j_2\rangle$, where $j_2\neq j_1-1$, $j_2\neq j_1+2$.
    \item $E_2=0$, existaions at NN sites are prohabited. Therefore the eigenstates are $|E_2^{(j_1j_2j_3)}\rangle = |j_1, j_2, j_3\rangle$, where $j_1 \neq j_2\pm 1$, $j_1\neq j_3\pm1$ and $j_3\neq j_2\pm1$.
\end{enumerate}

The projection operator to the three-excitation bound state subspace,
\begin{equation}
    P_0 = \sum_j |E_0^{(j)}\rangle\langle E_0^{(j)}|,
    \label{eq:3_part_proj0}
\end{equation}
and the projection operator to the orthogonal subspace
\begin{align}
    P_1 &= \sum_{j_1,J_2}\frac{1}{E_0-E_1}|E_1^{(j_1j_2)}\rangle\langle E_1^{(j_1j_2)}| \nonumber\\
    &+\sum_{j_1,J_2,j_3}\frac{1}{E_0-E_2}|E_2^{(j_1j_2j_3)}\rangle\langle E_2^{(j_1j_2j_3)}|.
    \label{eq:3_part_proj1}
\end{align}
The third-order effective Hamiltonian in terms of projection operators and perturbative Hamiltonian is given by,
\begin{align}
    \hat{H}_{eff}^{(3)} &= E_0P_0 + P_0\hat{H}_1P_0+P_0\hat{H}_1P_1\hat{H}_1P_0 \nonumber\\
    &+P_0\hat{H}_1P_1\hat{H}P_1\hat{H}_1P_0 \nonumber\\
    &-\frac{1}{2}(P_0\hat{H}_1P_0\hat{H}_1P_1^2\hat{H}_1P_0+P_0\hat{H}_1P_1^2\hat{H}_1P_0\hat{H}_1P_0).
    \label{eq:3_part_pur_exp}
\end{align}
Substituting Eq.~\ref{eq:pert_ham}, ~\ref{eq:3_part_proj0} and \ref{eq:3_part_proj1} into Eq.~\ref{eq:3_part_pur_exp} we get the following effective Hamiltonian
\begin{eqnarray}
        \hat{H}_{eff}^{(3)} &=& \frac{-J_1}{J_z}\left(-2J_2+\frac{J_1^2}{J_z}+\frac{4J_2^2}{J_z}\right)\sum_{i=1}^{L-3}(\hat{B}_i^\dagger\hat{B}_{i+1}+\text{H.c.}) \nonumber\\
        &+& \frac{J_2^3}{J_z^2}\sum_{i=1}^{L-4}(\hat{B}_i^\dagger\hat{B}_{i+2}+\text{H.c.})\nonumber\\
        &+& \left(\frac{2J_1^2}{J_z}+\frac{4J_2^2}{J_z}-\frac{6J_1^2J_2}{J_z^2}\right)\sum_{i=1}^{L-2}\hat{B}_i^\dagger\hat{B}_{i},
        \label{eq:3part_eff_ham}
\end{eqnarray}
where $B_i=\hat{\sigma}_{i-1}^-\hat{\sigma}_i^-\hat{\sigma}_{i+1}^-$. This effective Hamiltonian shows a good agreement with our numerical results.

\end{document}